\documentclass[aps,pra,reprint]{revtex4-1}
\usepackage{graphicx}
\usepackage{amsmath, amsthm, amssymb}
\usepackage{multirow}
\usepackage{subfigure}
\usepackage{bm}
\usepackage{xspace}            

\begin{document}

\newcommand{\tab}{Table}
\newcommand{\sect}{Sec.}
\newcommand{\sectf}{Section}
\newcommand{\sects}{Sections}
\newcommand{\sectsf}{Sections}
\newcommand{\fig}{Fig.}
\newcommand{\figf}{Figure}
\newcommand{\figs}{Fig.s}
\newcommand{\figsf}{Figures}
\newcommand{\eq}{Eq.}
\newcommand{\eqf}{Equation}
\newcommand{\eqs}{Eq.s}
\newcommand{\eqsf}{Equations}
\newcommand{\degrees}{\ensuremath{^\circ}}
\newcommand{\IM}{\operatorname{Im}}
\newcommand{\RE}{\operatorname{Re}}
\newcommand{\Tr}{\operatorname{Tr}}
\newcommand{\rb}{$^{85}$Rb}
\newcommand{\linplin}{\ensuremath{\rm lin\perp lin}\xspace}

\title{Stimulated Raman Adiabatic Passage for Improved Performance of a Cold Atom Electron and Ion Source}

\newcommand{\unimelb}{School of Physics, The University of Melbourne, VIC 3010, Australia}

\author{B. M. Sparkes}
\author{D. Murphy}
\author{R. J. Taylor}
\author{R. W. Speirs}
\author{A. J. McCulloch}
\author{R. E. Scholten}
\email[]{scholten@unimelb.edu.au}

\affiliation{\unimelb}
\date{\today}

\begin{abstract}
We implement high-efficiency coherent excitation to a Rydberg state using stimulated Raman adiabatic passage in a cold atom electron and ion source. We  achieve an efficiency of 60\% averaged over the laser excitation volume with a peak efficiency of 82\%, a 1.6 times improvement relative to incoherent pulsed-laser excitation. Using pulsed electric field ionization of the Rydberg atoms we create electron bunches with durations of 250\,ps. High-efficiency excitation will increase source brightness, crucial for ultrafast electron diffraction experiments, and coherent excitation to high-lying Rydberg states could allow for the reduction of internal bunch heating and the creation of a high-speed single ion source.
\end{abstract}


\maketitle

\section{Introduction}
\label{stirap:intro}

Cold atom electron and ion sources (CAEIS) \cite{Claessens2005,Claessens2007,McCulloch2011,Engelen2013,Hanssen2008,Steele2010}, based on the photo-ionization of laser-cooled gases, offer the potential for dramatic improvements for electron diffraction, nanofabrication and microscopy. One of the main drivers for the development of a CAEIS is the long-term goal of creating ``molecular movies'': to probe dynamic processes with atomic spatial and temporal resolution.  Substantial advances towards this goal have been demonstrated with electron  \cite{Dwyer2006, Siwick2003, Harb2008, Sciaini2009, Tokita2009, VanOudheusden2010, Ishikawa2015} and X-ray \cite{Chapman2006, Chapman2011, Seibert2011, Boutet2012, Kimura2014, Kupitz2014, Nogly2014} single-shot ultrafast diffraction.

A key metric for ultrafast diffraction is the normalized beam brightness \cite{Luiten2007}. Conventional electron sources are not sufficiently bright for collecting single-shot diffraction signals from weakly scattering molecules or nanocrystals. Beam brightness is proportional to particle flux, which for a CAEIS depends linearly on the density of the cold-atom cloud and the photo-ionization probability, or efficiency. To date, most CAEIS experiments have used photo-excitation with pulsed lasers in the presence of a static ionizing electric field. The incoherent nature of the excitation has limited the peak efficiency to 50\%, while requiring high laser power due to saturation of the conventional excitation process.

\begin{figure}
\centering
\includegraphics[width=\hsize]{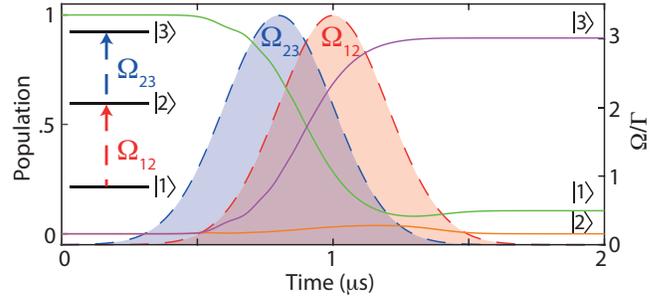}
\caption{Simulation of high efficiency excitation using stimulated Raman adiabatic passage in a three-level ladder system. Solid lines represent atomic state populations (left-hand axis), dashed and filled lines represent Rabi frequencies $\Omega$ normalized to the intermediate state decay rate $\Gamma$ (right-hand axis).}
\label{stirapfig:stirap}
\end{figure}

Stimulated Raman adiabatic passage (STIRAP) \cite{Bergmann1998} offers a mechanism for increasing the CAEIS excitation efficiency, particular in an optically dense cold atom target, and therefore improving source brightness. Here we are specifically interested in excitation to Rydberg states of rubidium-85 in a three-level ladder system (Fig.~\ref{stirapfig:stirap}) \cite{Low2012}. By first illuminating the atoms with light of a frequency $\omega_{23}$, resonant with the $\left | 2 \right\rangle \rightarrow \left | 3 \right\rangle$ transition, and then a second temporally over-lapping light field of frequency $\omega_{12}$, a dark state is formed by a coherent superposition of states $\left | 1 \right\rangle$ and $\left| 3 \right\rangle$. As the intensity of the light fields change, the atomic state transitions from state $\left| 1 \right\rangle$ to $\left| 3 \right\rangle$, bypassing $\left| 2 \right\rangle$. Figure~\ref{stirapfig:stirap} shows the population of the three states during the above-mentioned ``counter-intuitive'' pulse sequence, simulated using optical Bloch equations for a ladder system \cite{Sevincli2011a} with Rabi frequencies $\Omega_{12}$ and $\Omega_{23}$.  

STIRAP is a robust technique and, provided the adiabatic condition is met ($\Omega_{\mathrm{eff}}\, \tau>10$, where $\Omega_{\mathrm{eff}}=\sqrt{\Omega_{12}^2+\Omega_{23}^2}$ is the effective Rabi frequency and $\tau$ is the interaction time), high efficiency excitation is possible with a variety of different individual Rabi frequencies, pulse delays and shapes. Experiments to date have demonstrated peak excitation efficiencies up to 90\% \cite{Cubel2005, Deiglmayr2006, Takekoshi2014a}, which would increase the brightness of a CAEIS by a factor of 1.8.

STIRAP also enables a method for producing very short bunches, and therefore for observing atomic-scale dynamics, by following excitation with pulsed-electric-field ionization \cite{Taban2008}. This method will lead to a longitudinal compression of the bunch following ionization: the electrons liberated at later times will be accelerated by a larger field, allowing for ultra-short bunches at the sample without ultra-high electron densities, and therefore large Coulomb-driven expansion, at the source. Rydberg states have long lifetimes (tens to hundreds of microseconds) and relatively low ionization thresholds ($\rm 600\,V\,cm^{-1}$ for $\rm 30 S_{1/2}$), easing experimental demands on the pulsed electric field supply. The coupling strength of Rydberg transitions is much higher in the absence of an electric field, so that much lower laser power is required with a pulsed electric field compared to excitation in a static field, making STIRAP excitation a viable option. Combining STIRAP excitation and fast pulsed-field ionization has the potential to create bunches that are cold, bright and ultrafast, which is difficult to replicate with incoherent ultrafast laser ionization \cite{Engelen2013,McCulloch2013}. 

The large dipole moments of Rydberg atoms enables Rydberg blockade, where excitation of one atom inhibits the excitation of other atoms close by \cite{Sevincli2011a,Low2012}. Rydberg blockade can, in principle, reduce disorder-induced heating \cite{Murillo2001,Kuzmin2002}, and thereby reduce emittance and increase focusabiltiy in a CAEIS \cite{Murphy2015}. By enforcing a  separation between Rydberg atoms larger than the laser excitation volume, blockade can allow selective excitation of discrete separated atoms, and thereby create a deterministic single ion source \cite{Saffman2010a,Beterov2011,Ates2013}.

With the much-reduced laser power required, STIRAP can also be used for high-efficiency continuous operation, with increased average current relative to pulsed trap-based CAEISs \cite{Kime2013,Knuffman2013,TenHaaf2014,Wouters2014}. Continuous sources are preferred for sub-nanometer ion beam milling, imaging and doping in semiconductor device fabrication. A continuous source of cold ions has recently been demonstrated using Rydberg excitation with a current of up to 130\,pA \cite{Viteau2016}, a 40-fold increase over direct, above-threshold ionization methods, illustrating the advantage of coherent excitation methods.

Here we present a CAEIS based on STIRAP excitation in a magneto-optical trap (MOT), with a volume-averaged excitation efficiency of 60\% and a corresponding peak efficiency of 82\%, 1.6 times the maximum possible with direct excitation.  We also use a streak method to investigate the temporal profile of the bunches created via electric-field ionization, and finally we discuss how STIRAP could be implemented in an atomic beam-based CAEIS.

\section{Method}
\label{stirap:method}
\begin{figure}[h]
\centering
\includegraphics[width=\hsize]{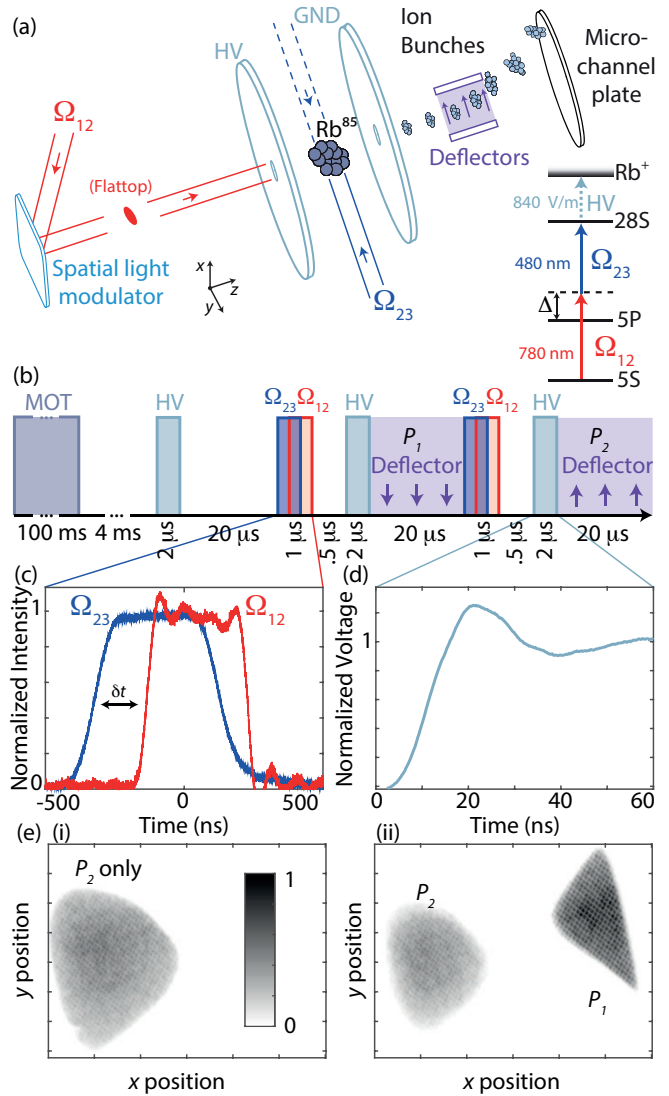}
\caption{(a) Cold atom ion source: HV refers to the high-voltage supplied to the front accelerator plate; GND is the grounded plate; and $\Omega_{12}$ and $\Omega_{23}$ refer to the two STIRAP fields. Inset shows the level structure of Rb$^{85}$ used here, including the electric field ionization strength required and the one-photon detuning $\Delta$. (b) Timing sequence for STIRAP excitation, field-ionization and two-pulse measurements, using electrostatic deflectors to spatially separate the two pulses ($P_1$ and $P_2$). (c) STIRAP pulse sequence, with temporal separation $\delta t< 0$. (d) Time-dependence of front accelerator potential,  for $\rm V_{\rm max} = 100\,V$. (e) Example MCP images showing (i) just pulse two ($P_2$) and (ii) both pulses. Color bar in (i) shows scaling used for both MCP images.}
\label{stirapfig:method}
\end{figure}

The CAEIS set-up is based around a MOT of rubidium-85 atoms located between two accelerator electrodes, as described in previous work 
\cite{McCulloch2011, McCulloch2013} and shown in Fig.~\ref{stirapfig:method}(a). A typical experimental sequence is shown in Fig.~\ref{stirapfig:method}(b), starting with the MOT being loaded for approximately 100\,ms. After this time all laser and magnetic fields are switched off and allowed to decay for 4\,ms to ensure a field-free excitation region. The atomic density after 4\,ms of expansion was measured to be $\rho_a = 5 \times 10^9\,\rm atoms\,cm^{-3}$ using absorption imaging.

In contrast to previous CAEIS experiments, which used a large-bandwith pulsed 480\,nm blue laser for direct ionization via a Stark-shifted manifold \cite{McCulloch2011,Saliba2012,Murphy2014a,Speirs2015}, here we used a frequency-doubled and amplified 960\,nm laser diode. The continuous laser provided a high-power (300\,mW), narrow-linewidth ($<$500\,kHz) source of 480\,nm light to couple the intermediate $\rm 5P_{3/2}$ state to a Rydberg level ($\rm 28S_{1/2}$). The frequency was stabilized using an ultrastable optical reference cavity. 

The STIRAP process (see level structure, Fig.~\ref{stirapfig:method}(a)) was driven by an infrared 780\,nm narrow-linewidth ($200$\,kHz) diode laser with 60\,nW of power and a frequency 27\,MHz blue-detuned from the $\rm 5S_{1/2} \rightarrow 5P_{3/2}$ transition to reduce incoherent absorption by atoms outside the interaction volume. The continuous blue laser was red-detuned 27\,MHz from the $\rm 5P_{3/2} \rightarrow 28S_{1/2}$ transition. We define the one-photon detuning as $\Delta = +27\rm\,MHz$.

Temporal control of the excitation fields was achieved via double-pass acousto-optic modulators. Rectangular pulses were used, as illustrated in Fig.~\ref{stirapfig:method}(c), and we define the delay between the pulses $\delta t$ to be negative if the blue pulse started before the red. The excitation region was determined by the spatial overlap of the two laser beams. The spatial profile of the infrared laser beam, controlled via a spatial-light modulator, was a uniform circular cross section with a radius of $R_r=150\rm\,\mu m$ in the plane perpendicular to the direction of charged particle propagation. The blue laser beam was focused to a ribbon with Gaussian standard deviations of approximately $\sigma_{x}=150\rm\,\mu m$ by $\sigma_z=20\rm\,\mu m$ in the perpendicular and longitudinal directions respectively. The optical excitation was driven without an external electric field to avoid Stark-splitting and loss of coupling strength.  A potential difference was then applied to the electrodes, with a rise time of 4\,ns (Fig.~\ref{stirapfig:method}(d)). The threshold electric field strength required for ionization of the $\rm 28S_{1/2}$ is $840\rm\,V\,cm^{-1}$. Typically an accelerator field of $\rm 1400\,kV\,cm^{-1}$ was applied to ensure complete ionization. The liberated electrons or ions (depending on the polarity of the electric field) propagated  70\,cm before detection with a micro-channel plate (MCP) combined with a phosphor screen and CCD camera.

STIRAP was performed twice in quick succession using ion bunches to determine the ionization efficiency. The total charge in the first and second bunches, $N_1$ and $N_2$ respectively, are related to the efficiency $\mathcal{E}(x,z)$ by
\begin{eqnarray}
N_1 & \propto &\iiint\limits_V \mathcal{E}\left(x,z\right) \mathrm{d}x\, \mathrm{d}y\, \mathrm{d}z\\
N_2 & \propto &\iiint\limits_V \mathcal{E}\left(x,z\right)\left[1-\mathcal{E}\left(x,z\right) \right] \mathrm{d}x\, \mathrm{d}y \,\mathrm{d}z,
\label{stirapeq:spatialdependence}
\end{eqnarray}
where the spatial dependence of $\mathcal{E}(x,z)$ comes from the intensity profile of the blue laser (the product of two independent Gaussians in $x$ and $z$), and the interaction volume $V$ is bounded by the size of the infrared laser ($x^2+y^2=R_r^2$). The total volume-averaged efficiency  can be determined from the overall charge present, 
\begin{equation}
\mathcal{E}_{\mathrm{int}} = 1-\frac{N_2}{N_1}.
\label{stirapeq:efficiency}
\end{equation}
This two-pulse method therefore provides a measure of efficiency that is independent of the atomic density, excitation volume and MCP efficiency \cite{Cubel2005,Deiglmayr2006} if we assume minimal atomic movement inside the MOT between the two STIRAP events. 

$N_{1,2}$ are determined by area integration of the MCP images for pulses $P_{1,2}$ shown in Fig.~\ref{stirapfig:method}(e).  The phosphor screen on the MCP detector has a decay time on the order of milliseconds, too long to be able to temporally separate the signals from the two pulses. Instead, a deflector was used to spatially separate the two bunches. We used a variant on the two-pulse method to remove dependence on the MCP sensitivity, which is not perfectly uniform across the detector.  Measurements were made with just the second pulse to give $N_1$ (Fig.~\ref{stirapfig:method}(e)(i)), and then at the same location with both pulses spatially separated to determine $N_2$. 

\section{Results and Analysis}
\label{stirap:results}

\begin{figure}
\centering
\includegraphics[width=\hsize]{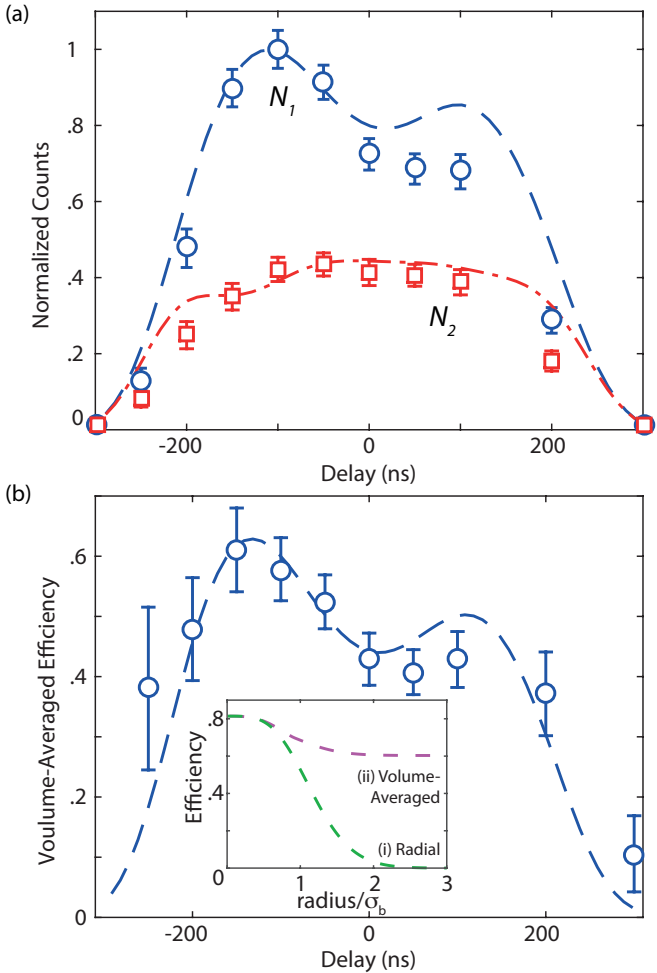}
\caption{(a) MCP Counts in first pulse ($N_1$ - blue, circles) and second pulse ($N_2$ - red, squares) as a function of the relative delay between the two excitation fields. Points indicate experimental data, with error bars determined from the standard deviation of 100~images, and lines indicate simulations using 200\,ns flattop pulses, peak Rabi frequencies $\Omega_{12} = \Omega_{23} = 15\rm\,MHz$, laser linewidths $\Gamma_{12}=\Gamma_{23}=500$\,kHz,  and $\Delta = 27\rm\,MHz$. (b) Efficiency calculated from the ratio of $N_2$ to $N_1$ using Eq.~\ref{stirapeq:efficiency}. Points indicate experimental data and lines indicate simulation. The inset shows (i) the calculated radial efficiency and (ii) the volume-averaged efficiency as a function of the blue laser beam radius, normalized to the Gaussian $\sigma_b$, at the optimal delay $\delta t = -150$\,ns.} 
\label{stirapfig:higheff}
\end{figure}

\subsection{STIRAP Efficiency}
\label{stirapsec:efficiency}
Figure~\ref{stirapfig:higheff}(a) shows the total integrated counts as a function of the delay between the pulses $\delta t$. Figure~\ref{stirapfig:higheff}(b) shows the volume-averaged efficiency calculated from the relative signals using Eq.~\ref{stirapeq:efficiency}, with the characteristic high efficiency seen when $\delta t <0$ (maximum of 60\% at $\delta_t = - 150$\,ns). 

Simulations were performed using optical Bloch equations \cite{Sevincli2011a} with experimentally realistic parameters (peak Rabi frequencies $\Omega_{12} = \Omega_{23} = 15\rm\,MHz$, $\Delta = 27\rm\,MHz$, intermediate state decay rate $\Gamma = 6$\,MHz, laser linewidths $\gamma_{12}=\gamma_{23}=500$\,kHz, for 200\,ns rectangular pulses with 100\,ns linear rise and fall times). Inset (i) of Fig.~\ref{stirapfig:higheff}(b) shows the simulated radial efficiency $\mathcal{E}\left[r\right]$ for a blue laser beam with Gaussian electric field profile with an arbitrary $1/e$ width of $\sigma_b$. Inset (ii) shows the volume-averaged efficiency $\int_0^r \mathcal{E}[r']dr'$ as the radius of integration increases to $\pm r$ in $z$ and either $\pm r$ or $\pm R_r$ in $x$, whichever is smaller. In the inset we have scaled $\sigma_x = \sigma_z=\sigma_b$ for simplicity, and used the fact that $R_r = \sigma_b$ . These simulations of the volume-averaged efficiency agree well with the experimental data in Fig.~\ref{stirapfig:higheff}. We can therefore infer a peak efficiency for STIRAP in the CAEIS of 82\% at the maximum blue intensity. Increasing the blue power would increase the maximum efficiency obtainable. However, with increased intensity comes the possibility of adding random phase and amplitude noise, which can limit the maximum efficiency obtainable \cite{Yatsenko2014}. Even without increasing the maximum intensity, for a uniform blue laser profile with intensity such that the Rabi frequency is the same as at the peak of our Gaussian profile, then we expect both volume-averaged and peak efficiencies would be 82\%. Non-uniform electric fields within the accelerator region, for example caused by charged particle accumulation on the electrodes, will also reduce the coupling strength, broaden the two-photon transition, and reduce the maximum efficiency.

The experimental results show a distinct reduction in signal compared to simulations for $\delta t >0$. This reduction is the opposite of the increase in signal seen elsewhere \cite{Cubel2005,Deiglmayr2006}, which was attributed to radiation trapping and Rydberg-Rydberg interactions. We use a large one-photon detuning to avoid absorption of the infrared laser outside the interaction zone. Any background absorption will lead to a large two-photon detuning for the re-radiated light interacting with the off-resonance blue light, causing a reduction in the excitation probability.  The accompanying optical pumping of the background atoms into the lower ground state during the first excitation/ionization event will reduce the fraction of re-radiating atoms for the second event, resulting mainly in a reduction of first pulse counts and, therefore, a reduction in the calculated efficiency. 

\subsection{Incoherent Excitation Efficiency}
\label{stirapsec:pbeff}

\begin{figure}
\centering
\includegraphics[width=.924\hsize]{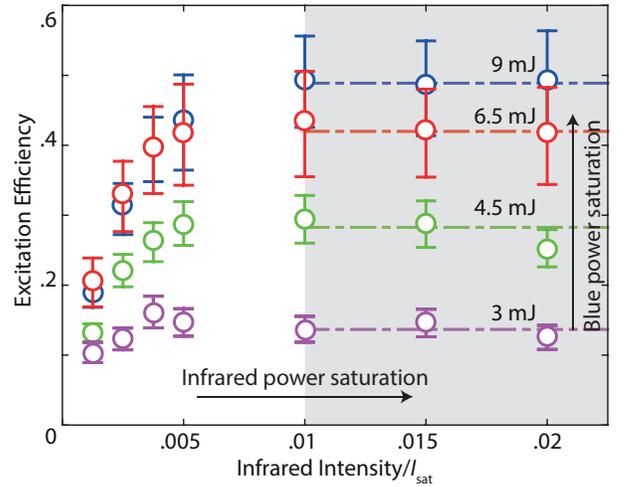}
\caption{Pulsed blue ionization efficiency as a function of infrared laser intensity normalized to saturation intensity ($I_{\mathrm{sat}}$) for different blue pulse energies. Numbers indicate the power of the pulsed blue laser, dashed lines show the saturation of ionization efficiency and shaded region denotes the region where the intermediate state becomes saturated.}
\label{stirapfig:pulsed}
\end{figure}

To quantify the improvement to CAEIS brightness provided by STIRAP, we measured the efficiency of pulsed 480~nm laser ionization using a variant of the two-pulse efficiency method. The pulsed and continuous blue laser beams were overlapped in counter-propagating directions (dashed lines in Fig.~\ref{stirapfig:method}(a)), perpendicular to the direction of charged particle propagation. The same infrared laser was used for both excitation processes, though the power and detuning were optimized separately for each: on resonance for pulsed-laser excitation and 27\,MHz detuned for STIRAP excitation. The accelerator field was applied before pulsed-laser excitation to reproduce ``normal'' ionization conditions for a CAEIS. $N_1$ was still defined as the signal for a single STIRAP pulse sequence, and $N_2$ as the signal for STIRAP excitation following excitation by the pulsed laser. Using this method, the efficiency of the pulsed blue laser as a function of infrared laser intensity and pulsed blue power was measured (Fig.~\ref{stirapfig:pulsed}).

The efficiency approaches 50\%, the maximum efficiency for incoherent excitation in a two-level system, as infrared laser intensity and pulsed blue energy increase. This limit arrises as the blue pulse duration (of order a few nanoseconds) is much faster than the infrared pumping rate, and so the intermediate state will not be refilled on the ionization timescale. Comparing the peak STIRAP excitation to this incoherent excitation peak gives an increase in efficiency by a factor of 60\%.

\subsection{Temporal Profile}
\label{stirapsec:temp}

\begin{figure}
\centering
\includegraphics[width=\hsize]{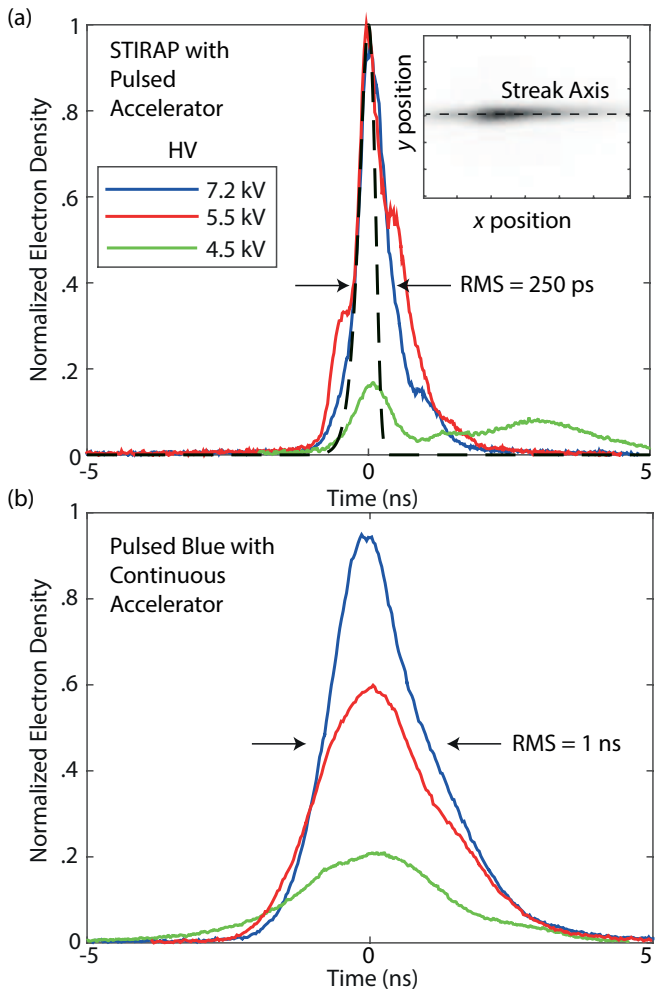}
\caption{Streak measurements of electrons created with: (a) STIRAP bunches with pulsed accelerator; and (b) pulsed blue laser with continuous accelerator at different accelerator potential differences HV. Inset shows a false-color streaked electron bunch as measured by the MCP. All traces are normalized to the same peak value. Solid lines indicate experimental data, dashed line indicates theory for hydrogenic ``red'' state with field switching behavior from Fig.~\ref{stirapfig:method}(d), normalized to height of experimental traces.}
\label{stirapfig:streak}
\end{figure}

The duration of the electron/ion bunches is an important parameter for most applications of a CAEIS. Coulomb-driven spatial expansion of charged bunches leads to temporal expansion, but the expansion is not significant for electrons because the propagation time from bunch creation to detection is too short.  Hence we investigated the temporal bunch shape using a streak method.  The electron bunches propagated through deflectors with a rapidly varying transverse potential, causing the bunch to ``streak'' across the detector, with the position of an electron on the detector being dependent on the time at which it entered the deflector region. The temporal profile of the bunch was then determined from a line profile along the streak, calibrated to the known geometry and time-varying potential difference. The streak measurements are shown in Fig.~\ref{stirapfig:streak} for bunches created with (a) STIRAP excitation followed by pulsed electric field ionization, and (b) pulsed blue ionization in a constant electric field.

For accelerator fields close to the electric-field ionization threshold of the $\rm 28S_{1/2}$ state, a broad secondary peak in the electron temporal distribution can be seen for the STIRAP bunches. This peak could be due to blackbody collisions transferring some atoms to lower energy states with a higher threshold ionization voltage \cite{Wang2007a}. The appearance of a much narrower secondary peak in both the 4.5\,kV and 5.5\,kV results also supports this explanation. Another possibility is non-ideal behavior of the high-voltage switch, for example by fast oscillations in the rising voltage.

The relative pulse heights show that a near-threshold voltage leads to only a small fraction of excited atoms being ionized. Once above the threshold voltage, this fraction approaches one, verified by the detection of only a very weak signal when performing a second electric field ionization pulse after a single STIRAP excitation sequence. The root mean square (RMS) duration of the STIRAP bunches, determined from the streak measurements of Fig.~\ref{stirapfig:streak}(a), was 250\,ps, varying only slightly for different accelerator potentials.

With an accelerator rise time on the order of nanoseconds, ionization will be diabatic (hydrogenic). Modelling an accelerator profile on Fig.~\ref{stirapfig:method}(d), the ionization rate for a ``red'' state of hydrogen (where Rydberg quantum numbers $m=n_1=0$, $n_2=n-1$) \cite{Hoe1981,Damburg1979} gives an RMS pulse width of 170\,ps (Fig.~\ref{stirapfig:streak}(a)), consistent with the initial rise in electron charge seen in the data of Fig.~\ref{stirapfig:streak}(a).

The measured duration of bunches produced with STIRAP excitation and field ionization compares favorably with that for pulsed blue excitation.  The bunch duration for incoherent excitation is determined by the temporal profile of the pulsed laser, which has a quoted total pulse length of 5\,ns, and produces bunches with duration of order 1\,ns RMS.  Ultrafast electron diffraction requires sub-picosecond pulses. With accelerator potentials of 30\,kV and 30\,ns electric field rise times, it has been shown that a bunch length of 80\,ps can be achieved \cite{Taban2008}.  To reduce the bunch duration below 1\,ps following STIRAP excitation, the maximum accelerator voltage would need to increase by an order of magnitude, and the switching time reduce to less than 1\,ns \cite{Claessens2005}. Achieving such electric field switching requires careful design of the MOT chamber and accelerator to avoid electrical discharge \cite{Taban2008}, and a very fast high-voltage switch, potentially using laser-triggered spark gap technology \cite{Brussaard2007}.  Alternately, an RF bunch compressor could be used \cite{VanOudheusden2010}.

\subsection{Robustness}
\label{stirapsec:robustness}

\begin{figure}
\centering
\includegraphics[width=\hsize]{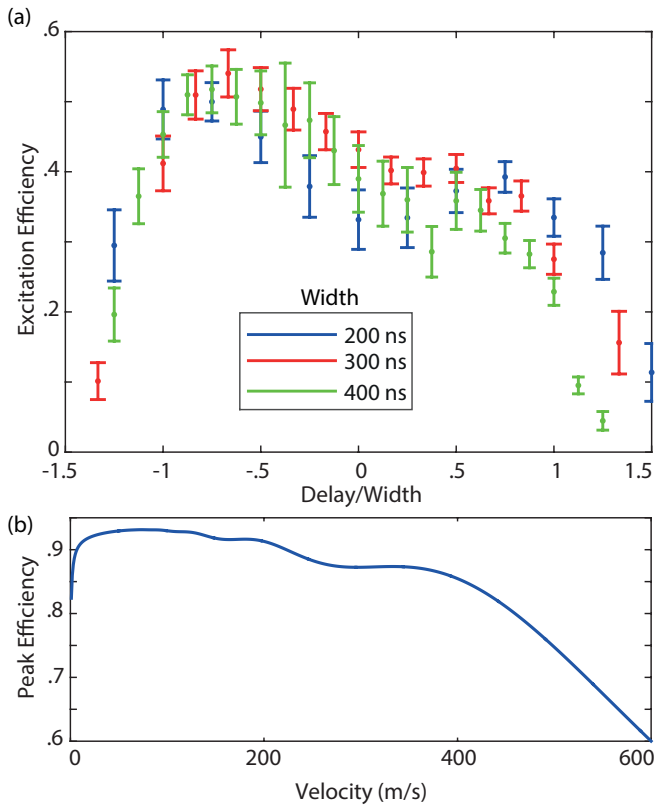}
\caption{(a) Efficiency as a function of the ratio of pulse delay $\delta t$ to flattop pulse width $w$. (b) Simulated excitation efficiency of STIRAP for an cold atom-beam source, as a function of atomic velocity with $\Omega_{12}=\Omega_{23}=15\rm\,MHz$, $\sigma_z = 15\rm\,\mu m$ and $\delta z = - \sigma_z$.}
\label{stirapfig:length}
\end{figure}

The effect of different STIRAP pulse widths $w$ was investigated (Fig.~\ref{stirapfig:length}(a)). The robustness of STIRAP excitation is apparent, since a difference in width by a factor of two has very little impact on either the maximum efficiency (50 to 55\%), or the time at which this occurs ($\delta t/w = -.75$ for the rectangular pulses used).

The robustness of STIRAP makes it ideally suited to next-generation cold atom ion sources based on atomic beams \cite{Kime2013,Knuffman2013,TenHaaf2014,Wouters2014}. The experimental situation described above, where atoms are stationary and the optical and electric fields are dynamic, is equivalent to an atomic beam system with atoms moving through spatially separated static optical fields and a region with an electric field gradient. The high temperature of the atoms along the direction of propagation will result in a large velocity spread. For instance, an experimentally practical atom beam temperature of 200$^\circ$C would lead to a most-probable velocity of $v_{zp}=305\rm\,m\,s^{-1}$ with standard deviation of $150\rm\,m\,s^{-1}$. The different velocities of the atoms are equivalent to a static atom seeing STIRAP fields with different temporal widths but a constant $\delta t/w$. Figure~\ref{stirapfig:length}(b) shows the peak efficiency calculated for such a system with Gaussian laser beam spatial profiles with $\sigma_z = 15$\,$\mu$m, and $\delta z = -\sigma_z$.  The efficiency remains above 80\% from 0 to $400\rm\,m\,s^{-1}$, so that a large proportion of the atomic population (66\%) will be excited with high efficiency.

High ion beam densities achieved using STIRAP excitation could lead to Coulomb explosion and a reduction in the focusability of the source. The density could be reduced by using Rydberg blockade with high principle quantum number $n\approx 100$ \cite{Low2012}. If the excitation volume is reduced to below one blockade radius, it will become possible to isolate separate ions spatially and temporally, to create a quasi-deterministic highly focusable single ion source with heralding provided by the liberated electrons \cite{Beterov2011,Ates2013}.

\section{Conclusions}
\label{stirap:conc}
We have shown that STIRAP can improve the excitation efficiency of a cold atom electron and ion source by a factor of 1.6, from a peak efficiency of 50\% with incoherent excitation, to 82\%. Further improvements are expected with higher laser power, greater uniformity of the electric field within the excitation region, and reduced phase noise in the excitation lasers.

We have also shown that STIRAP excitation and fast switching of the ionization electric field produces bunches with an RMS duration of 250\,ps.  Sub-picosecond bunches may be achievable with higher acceleration potentials and faster switching, and with an RF compressor, to satisfy the temporal criterion for imaging dynamic processes with atomic spatial and temporal resolution using ultrafast electron diffraction.

With continuous lasers and an atomic beam, STIRAP excitation will be directly applicable to next-generation continuous atom-beam based cold electron and ion sources. Finally, by using high efficiency STIRAP excitation to reach higher Rydberg states, the phenomena of Rydberg blockade could be used to create spatial ordering, and therefore reduce the temperature and increase the focusability of the bunches, as well as enabling a new approach to creating a deterministic single ion source.

\textbf{Acknowledgements}
BMS gratefully acknowledges the support of a University of Melbourne McKenzie Fellowship. This work was supported by the Australian Research Council Discovery Project DP140102102.
\bibliographystyle{unsrt}
\bibliography{library} 
\end{document}